\documentclass
[prl,twocolumn,a4paper,superscriptaddress,floatfix,showpacs,aps]{revtex4-1}%
\usepackage[english]{babel}
\usepackage[utf8]{inputenc}
\usepackage{amsmath}
\usepackage{natbib}
\usepackage{graphicx,epstopdf}
\usepackage{amssymb,epsfig,color,textcase}
\usepackage{hyperref}
\usepackage{amsfonts}
\usepackage{amssymb}
\usepackage{ulem}
\usepackage{graphicx}%
\providecommand{\U}[1]{\protect\rule{.1in}{.1in}}
\newcommand{\angstrom}{\textup{\AA}}

\begin {document}

\title {The role of temperature and Coulomb correlation in stabilization of CsCl-phase in FeS under pressure}

\author {A.~O.~Shorikov}
\affiliation{M.N. Miheev Institute of Metal Physics of Ural Branch of Russian Academy of Sciences, 18 S. Kovalevskaya Street, 620108 Yekaterinburg, Russia}
\affiliation{Ural Federal University, 19 Mira Street, 620002 Yekaterinburg, Russia}

\author {V.~V.~Roizen}
\affiliation{Moscow Institute of Physics and Technology, 9 Institutskiy per., Dolgoprudny, Moscow Region, 141701, Russia}

\author {A.~R.~Oganov}
\affiliation{Skolkovo Institute of Science and Technology, 3 Nobel Street, Moscow, 143026, Russia}
\affiliation{Moscow Institute of Physics and Technology, 9 Institutskiy per., Dolgoprudny, Moscow Region, 141701, Russia}

\author {V.~I.~Anisimov}
\affiliation{M.N. Miheev Institute of Metal Physics of Ural Branch of Russian Academy of Sciences, 18 S. Kovalevskaya Street, 620108 Yekaterinburg, Russia}
\affiliation{Ural Federal University, 19 Mira Street, 620002 Yekaterinburg, Russia}

\begin {abstract}
The iron-sulfur system is important for planetary interiors and is intensely studied, particularly for better understanding of the cores of Mars and Earth. Yet, there is a paradox about high-pressure stability of FeS: {\it ab initio} global optimization (at DFT level) predicts a $Pmmn$ phase (with a distorted rocksalt structure) to be stable at pressures above $\sim$~120 GPa, which has not yet been observed in the experiments that instead revealed a CsCl-type phase which, according to density functional calculations, should not be stable. Using quasiharmonic free energy calculations and the dynamical mean field theory, we show that this apparent discrepancy is removed by proper account of electron correlations and entropic effects.

\end {abstract}

\pacs {61.50.Ks, 62.50.-p, 64.70.Kb, 71.15.Mb, 71.27.+a}

\maketitle

The composition of planets core is a topic of intense research. The main components of Earth’s core, iron and nickel are mixed with a small amount of other light elements~\cite{Poirier1994}. However, the chemical composition and crystal structure of the core compounds is still the subject of discussion. Sulfur is seen as one of the most preferred candidates to be present in the core~\cite{Seagle2006} and Fe$_{1-x}$S is one of the most widespread sulfides on Earth (encountered also in lunar and meteoric samples~\cite{Fei1995, Kamimura1992, Kusaba1997, Taylor1970, Wang2005}). High-pressure behavior of iron sulfide (FeS) has been investigated through both high-pressure experiments and {\it ab initio} simulations in multiple previous studies~\cite{King1982,Sherman1995,Fei1995,Kusaba1997,Kobayashi1997,Alfe1998,Kusaba1998,Nelmes1999,Takele1999,Rueff1999,Vocadlo2000,Marshall2000,Kavner2000,Martin2001,Urakawa2004,Kobayashi2004,Ono2006,Ushakov2017}. FeS exhibits rich polymorphism and its magnetic properties and phase diagram under high pressure are of great interest in both condensed matter physics and planetary science, and have been extensively studied.

 Stoichiometric FeS has a NiAs-type (B8) related hexagonal structure (troilite, FeS I) at ambient condition with $P\bar{6}2c$ space group~\cite{Marshall2000}. The onset of a long-range magnetic order is observed at $T_N \sim 600$~K.  Previous experimental studies demonstrate a series of phase transitions with increasing pressure at room temperature; troilite transforms to a MnP-type structure (FeS II)  with the orthorhombic space group $Pnma$ above 3.4 GPa~\cite{King1982,Keller1990} and further to a monoclinic structure (FeS III) above 6.7 GPa.  This transition is accompanied by a lattice volume collapse~\cite{Kamimura1992} and a change in the crystal symmetry (space group $P2_1/a$). The structural change from FeS II to III involves abrupt breaking of the long-range magnetic order~\cite{King1978,Kobayashi1997,Marshall2000}, spin transition of iron, and metal-semiconductor transition. FeS IV (hexagonal structure) and FeS V (NiAs-type structure) are also known to exist at high pressure and high temperature.  A phase transition to FeS VI with $Pnma$ space group (MnP-type) was found to occur above 30~GPa and 1300~K~\cite{Ono2007}. {\it Ab initio} calculations at higher pressures predicted transformation from the monoclinic phase of FeS to the CsCl-type phase (B2) with $Pm\bar{3}m$ space group~\cite{Martin2001}. This result has been confirmed by experiment~\cite{Sata2008}. The CsCl-type phase was synthesized at 1300~K and 186~GPa~\footnote{The CsCl-phase was synthesized at 186 GPa after heating to 1300 K, and this P,T-point (shown by red star on our phase diagram) corresponds to stability of this phase. Then this phase was quenched to 298 K, where it probably is metastable.}. However, there is a contradiction with more recent band-structure calculations~\cite{Ono2008} which predicted another phase with $Pmmn$ symmetry to be stable, while the CsCl-type structure is metastable at zero Kelvin (by 0.1-0.15 eV/atom). It should be noted that the CsCl-type structure is stable at high pressures in other iron alloys, such as FeSi~\cite{Dobson2002}. 
 
Such discrepancy between theoretical and experimental results is quite intriguing. One can assume that the CsCl-phase is stabilized by thermal effects. The results of Gibbs free energy calculations within the quasi-harmonic approximation confirmed this hypothesis to a certain degree, while placing the CsCl-type structure’s stability field much higher in terms of pressure and temperature in comparison with the experimental data~\cite{Bazhanova2017}. Such difference cannot be put down to numerical errors, and that is why we assumed that Coulomb correlations too may be crucial in the stabilization of the CsCl-phase. To confirm this, we conducted calculations by combining the Generalized Gradient Corrected Local Density Approximation and the Dynamical Mean–Field Theory (DFT+DMFT). Phonon calculations were run using the finite displacements method and allowed us to take thermal effects into account. Combining the results of our computational modeling, we calculated Gibbs free energy and constructed a (P,T)-phase diagram.
\begin {figure}
\vspace{5mm}
\includegraphics [width=0.16\textwidth,clip=true]{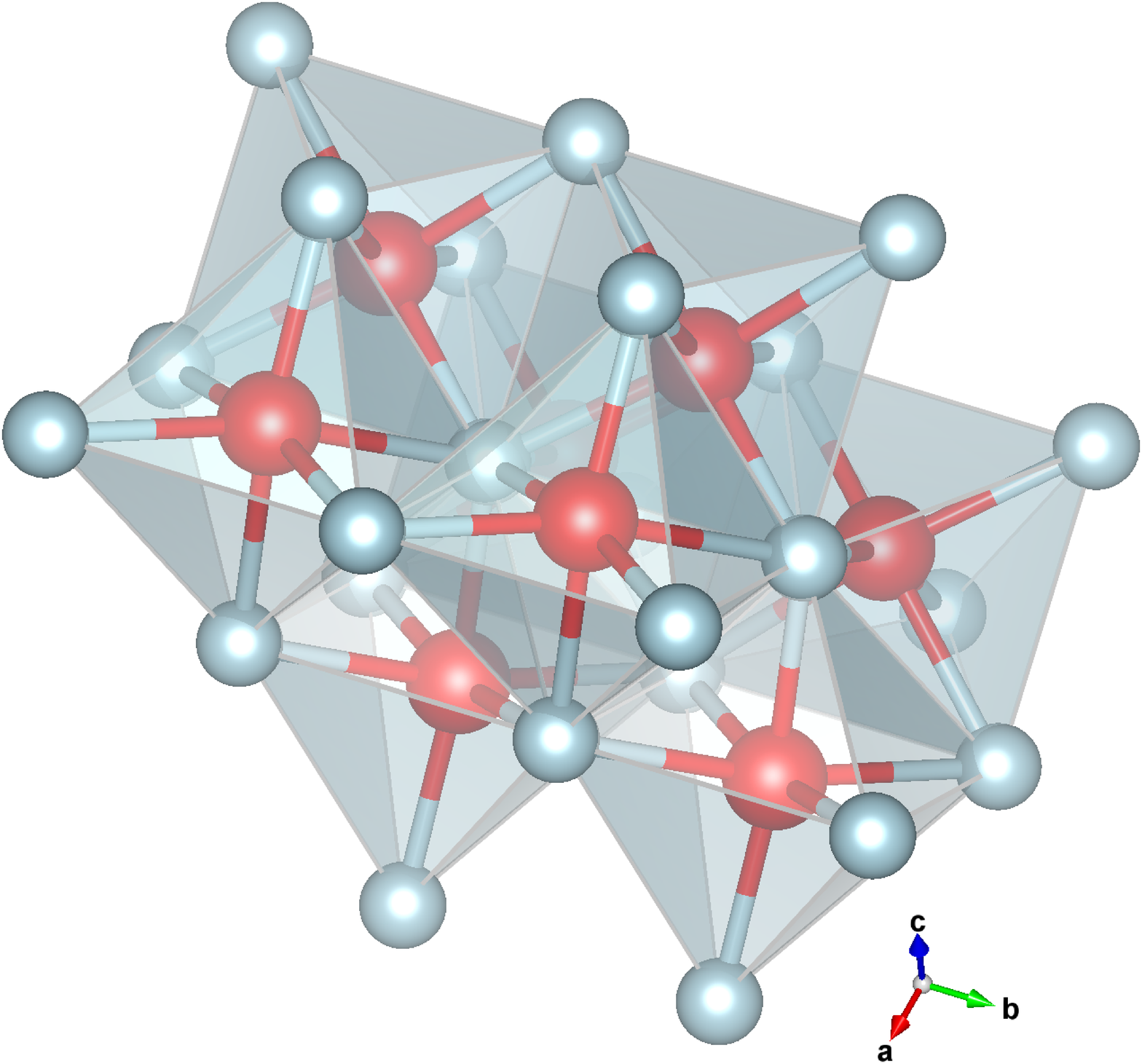}
\includegraphics [width=0.15\textwidth,clip=true]{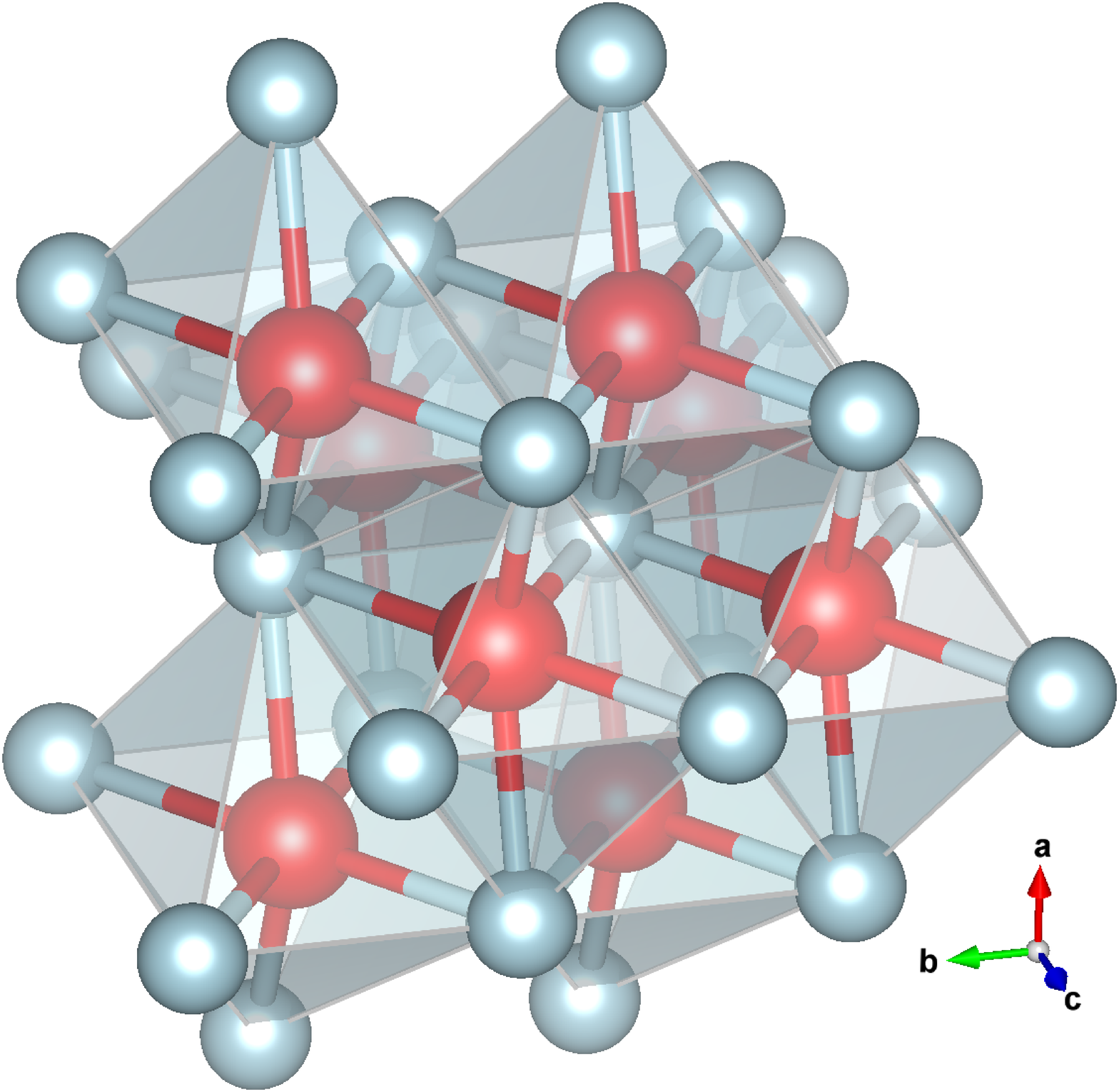}
\includegraphics [width=0.15\textwidth,clip=true]{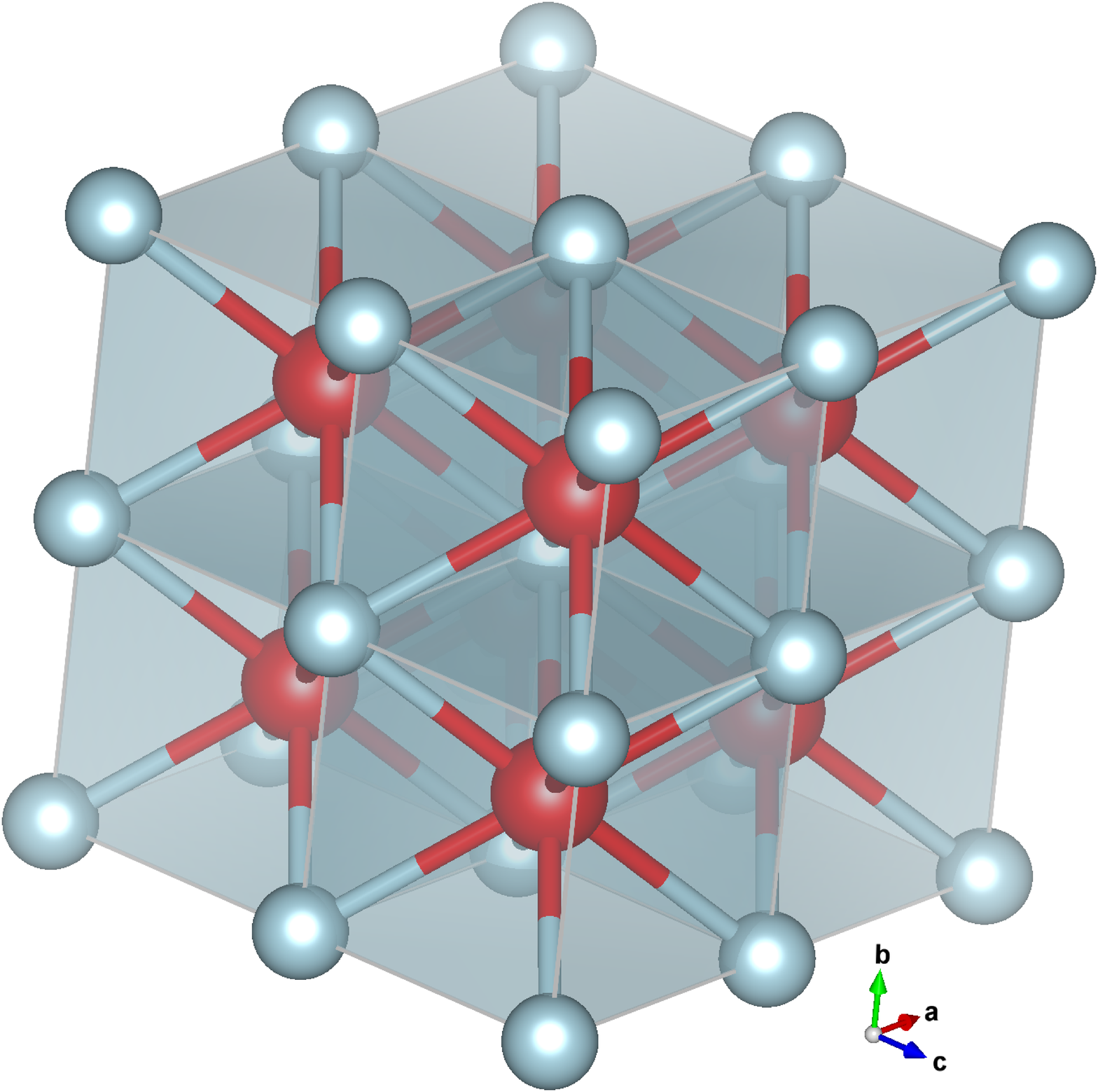}
\caption {(Color online). Crystal structures of  Pnma (left) Pmmn (middle) and CsCl (right) phases of FeS. Iron is shown with red and sulfur with blue balls. The crystal structures were drawn using VESTA~\cite{Momma2011}.}
\label {fig:cryst_str}
\end {figure}

As the first step the relaxation of atomic positions was applied to  three  crystal structures under investigations (namely $Pnma$, $Pmmn$ and $Pm\bar{3}m$) for a wide pressure range (-10:400 GPa) using the VASP code\cite{Kresse1996}. We used the exchange-correlation potential in the form proposed by Perdew, Burke, and Ernzerhof~\cite{Perdew1996}. PAW potentials with an [Ar] core (radius 2.3 a.u.) and [Ne] core (radius 1.9 a.u.) for Fe and S atoms, respectively, and a plane wave kinetic energy cut-off of 600 eV were used. Structure relaxations employed homogeneous-centered meshes with reciprocal space resolution of $2\pi \times 0.02 \angstrom^{-1}$ and Methfessel-Paxton electronic smearing with $\sigma$=0.16 eV. In order to take into account the correlation effects in the $d$-shell of iron we applied the so-called DFT+DMFT approach which exploits the advantages of two other methods widely used nowadays: the noninteracting band structure, $\varepsilon(\vec k)$, obtained within the density function theory (DFT) takes into account all the peculiarities of $\varepsilon(\vec k)$ for a given material, while the dynamical mean-field theory (DMFT) takes care of many-body effects such as Coulomb correlations~\cite{Anisimov1997,Held2006}. This method was successfully used in investigating different magnetic phenomena, including spin state transitions~\cite{Shorikov2010a,Kunes2008,Shorikov2015a,Skorikov2015}. In contrast to LDA+U or GGA+U approaches, it allows both considering frequency dependence of the self-energy and simulating a paramagnetic state.  The noninteracting GGA calculations were performed using the pseudo-potential method as implemented in Quantum ESPRESSO ~\cite{Giannozzi2009}. We used the Wannier function projection procedure~\cite{Korotin2008} to extract the noninteracting GGA Hamiltonian $H_{GGA}$ which included both Fe $3d$ and S $2p$ states. The full many-body Hamiltonian to be solved by the DFT+DMFT is written in the form:

\begin{equation}
\hat H= \hat H_{GGA}- \hat H_{dc}+\frac{1}{2}\sum_{i,m,m',\sigma,\sigma^{\prime}}
U^{\sigma,\sigma^{\prime}}_{m,m'}\hat n_{i,m,\sigma,}\hat n_{i,m',\sigma^{\prime}}.
\label{eq:ham}
\end{equation}
Here $U^{\sigma,\sigma^{\prime}}_{m,m'}$ is the Coulomb interaction matrix and  $\hat n_{im,\sigma}$ is the occupation number operator for the $d$ electrons with orbitals $m$ or $m'$ and spin indexes $\sigma$ or $\sigma^{\prime}$ on the $i$-th site.  The term $\hat H_{dc}$ stands for the {\it d}-{\it d} interaction  already accounted for in the DFT, the so called double-counting correction which was chosen to be $\hat H_{dc}=\bar{U}(n_{\rm dmft}-\frac{1}{2})\hat{I}$\cite{Anisimov1997}. Here $n_{\rm dmft}$ is the self-consistent total number of {\it d} electrons obtained within the DFT+DMFT and $\bar{U}$ is the average Coulomb parameter for the {\it d} shell. The elements of $U_{m,m'}^{\sigma\sigma'}$ matrix are parameterized by $U$ and $J_H$ according to the procedure described in Ref~\cite{Lichtenstein1998}. 

The effective impurity problem for the DMFT was solved by the hybridization  expansion Continuous-Time Quantum Monte-Carlo method (CT-QMC) \cite{Gull2011}. Calculations were performed for all the structures in the paramagnetic state at temperatures of 1160~K, 2000~K, 3000~K, 4000~K and 5000~K, using the AMULET code~\cite{AMULET}. For the sake of simplicity, we used the same set of Coulomb parameters for all the structures and pressures (unit cells) under investigation. The on-site Hubbard $U=6$ eV and Hund's intra-atomic exchange $J_H=0.95$ eV were estimated in QE using constrained GGA calculations~\cite{Anisimov2009a}. Note that these values agree well with the results of previous calculations of $U$ for other Fe sulfides and oxides at high pressure~\cite{Dyachenko2016,Ushakov2017} on the same Wannier functions which were applied  to  construct a small noninteracting Hamiltonian used in the subsequent DFT+DMFT calculations~\cite{Anisimov1997,AMULET}. 

Total energy was calculated within the DFT+DMFT as:
\begin{eqnarray}
E=E_{GGA}+\langle \hat{H}_{GGA}\rangle - \sum\limits_{m,k} \epsilon_{m,k}^{GGA} \nonumber \\
+\frac{1}{2}\sum\limits_{i,m,m',\sigma,\sigma'} U_{m,m'}^{\sigma,\sigma'}\langle\hat{n}_{i,m,\sigma} \hat{n}_{i,m',\sigma'}\rangle -E_{DC}
\label{e_tot}
\end{eqnarray}
Here $E_{GGA}$ stands for the total energy obtained within GGA. The third term on the right-hand side of Eq.~\ref{e_tot}  is the sum of the Fe-d, S-p valence state eigenvalues calculated as the thermal average of the GGA Wannier Hamiltonian with GGA Green function $\sum\limits_{m,k} \epsilon_{m,k}^{GGA} = \frac{1}{\beta}\sum\limits_{n,{\bf k}} Tr[H_{GGA}({\bf k})G_{{\bf k}}^{GGA}(i\omega_n)]e^{i\omega_n0^{+}}$. $\langle \hat{H}_{GGA}\rangle$ is evaluated in the same way but with the Green function which includes self-energy. The fourth term represents the interaction energy, here  $\langle\hat{n}_{i,m,\sigma} \hat{n}_{i,m',\sigma'}\rangle$  is the double occupancy matrix calculated in the DMFT. The double-counting correction $E_{DC} = \frac{1}{2}\sum\limits_{i,m,m',\sigma,\sigma'} U_{m,m'}^{\sigma,\sigma'}\langle\hat{n}_{i,m,\sigma}\rangle\langle \hat{n}_{i,m',\sigma'}\rangle$ corresponds to the average Coulomb repulsion between electrons in the Fe 3$d$ Wannier orbitals calculated from the self-consistently determined local occupancies. 

In order to evaluate pressure,  we fit our total energies to the third-order Birch-Murnaghan equation of states~\cite{Birch1947} separately for all the crystal structures under investigations and both the HS and the LS solutions obtained for the $Pnma$ one. The enthalpies ($H=E+PV$) of each phase were calculated to investigate the phase stability and transition pressures.

\begin {figure}
\vspace{5mm}
\includegraphics [width=0.425\textwidth,clip=true]{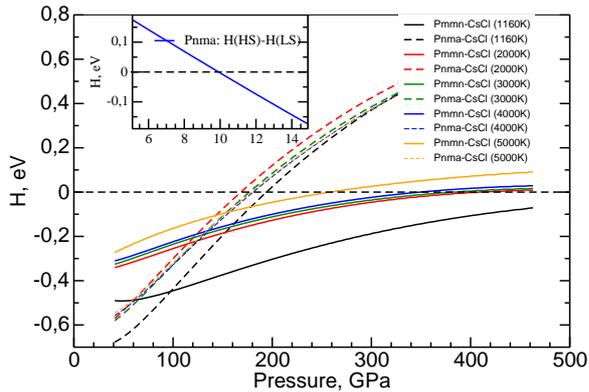}
\caption {(Color online) The difference of enthalpy for  Pmmn (solid lines) and Pnma (dotted lines) to CsCl phase (shown as zero line) in FeS as a function of pressure from DFT+DMFT calculations. Difference of enthalpy of HS to LS $Pnma$ phases  for T=1160~K is shown in the inset.}
\label {fig:enth_all}
\end {figure}

The enthalpies  calculated  from the DFT total energy as the first step showed that the CsCl-type phase is unstable, which agrees well with the previous study by Ono{\it et al.}~\cite{Ono2008}, and the transition pressure from MnP-type ($Pnma$) and $Pmmn$ phases is about 140 GPa. On-site Coulomb repulsion which we take into account within the DFT+DMFT method brings about a dramatic change in the results. Note that all the phases at high pressures were obtained in the low spin (LS) and paramagnetic state. The magnitude of average local moment $\langle m_z^2\rangle$ is different from zero and decreases gradually with pressure, e.g. it is $\sim$ 4 $\mu_B^2$ at ambient pressure (AP) and 1.8 $\mu_B^2$ at $\sim$ 400~GPa for the CsCl-type phase. Similar behavior was observed for the $Pnma$ phase (3.08 $\mu_B^2$ and 1.79 $\mu_B^2$) and $Pmmn$ phase (2.56$\mu_B^2$ and 1.87 $\mu_B^2$) at AP and $\sim$~400~GPa, respectively.  The calculated enthalpies with respect to those of the CsCl-type structure (which is shown as a zero line) are presented in Fig.~\ref{fig:enth_all} which shows that the CsCl-type can be stabilized by both pressure and temperature.  We believe the temperature plays an important role in $Pmmn$-CsCl-type transition. At 1160~K the transition pressure should lie beyond $\sim$500-600 GPa, while at T=2000~K the CsCl-type phase become stable at $\sim$410~GPa. We also calculated the critical pressure for HS to LS transition in $Pnma$ phase. The result is shown in the inset in Fig.~\ref{fig:enth_all}. The calculated transition pressure is about 10 GPa, which agrees fairly well with the experimental data (6.7 GPa at room temperature). Recently Ushakov{\it et al.}~\cite{Ushakov2017} have shown that a similar HS to LS transition in  troilite (FeS I, $P\bar{6}2c$ space group) can be reproduced by cell volume reducing. Though the correlation effects taken into account in the DFT+DMFT method allow us to reproduce transition into the CsCl-type structure observed experimentally, the calculated stability field lies beyond the experimental pressures and temperatures. There should be another mechanism, such as the vibration mechanism, capable of stabilizing the CsCl-structure.

In order to confirm this suggestion, we take into account thermal effects phonons and thermodynamic properties of FeS phases using the finite-displacement approach as implemented in the PHONOPY code~\cite{Togo2008,Togo2015}. To perform phonon calculations, all the structures were fully relaxed, with a cutoff of 600 eV and relaxation going on until all the force components dropped below 0.01 meV/\AA ~in absolute value. We constructed supercells (typically $2\times2\times2$, with dimensions of over 10~\AA) and displaced atoms by 0.01 \AA~  to obtain the forces which were then used to construct the force constants matrix. Then the dynamical matrix was constructed and diagonalized at a very dense reciprocal-space mesh.  
Next the results of the DFT+DMFT and phonon calculations were combined in order to compute Gibbs free energies and construct a phase diagram in (P,T)-coordinates.
\begin {figure}[ht!]
\vspace{5mm}
\includegraphics [width=0.425\textwidth,clip=true]{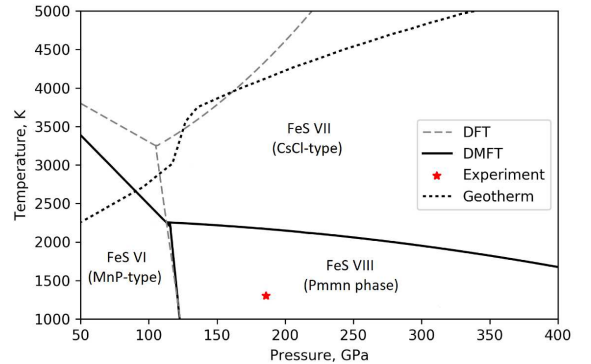}
\caption {(Color online) Phase diagram as calculated within the DFT and DFT+DMFT methods, including vibrational effects. The temperature profile of the Earth (geotherm) is shown for a reference. The experimental point of the CsCl-type structure is marked with a red star.}
\label {fig:pt}
\end {figure}

We addressed the task based on the following reasoning:
\begin{eqnarray}
F(V,T)=E_{elect}(V,T)+F_{vib}(V,T) 
\label{helmholtz}
\end{eqnarray}  
\begin{eqnarray}
E_{elect}(V,T)=E_{DFT}(V)+E_{DMFT}(V,T)
\label{elect}
\end{eqnarray}  
\begin{eqnarray}
F_{vib}(V,T)=E_{zp}(V)+\int\limits_0^T C_{V}dT+S_{vib}(V,T)*T
\label{vib}
\end{eqnarray}  
\begin{eqnarray}
P=-\frac{dF(V,T)}{dT} = P_{elect}(V,T)+P_{vib}(V,T)
\label{pressure}
\end{eqnarray} 
\begin{eqnarray}
G(P,T)=F(V,T) + P(V,T)*V
\label{gibbs}
\end{eqnarray} 
The equations of state were fitted using the Vinet equation of state ~\cite{Vinet1987} and a third-degree polynomial function. The phase diagram was plotted by calculating differences in the Gibbs free energies.

The phase diagram calculated within the DFT+DMFT method, including vibrational effects, is shown in  Fig.~\ref{fig:pt}. One can see from the diagram that the stability field of the CsCl-type phase in the DFT+DMFT approximation is sufficiently shifted to lower temperatures. These results lie much closer to the experimental conditions of the phase transition from the $Pnma$-phase to the CsCl-type (186~GPa and 1300~K). However, the experimental points still lie in the computed stability field of the $Pmmn$-phase. This discrepancy, albeit insignificant in this case, can be corrected by moderately adjusting the on-site Hubbard $U$ (which  was fixed in the present study, but should decrease slightly with pressure due to more effective screening) and Hund's intra-atomic exchange $J_H$. The diagram shows that the correlation effects have a strong impact mainly on the stability field of the CsCl-type phase, leaving other transitions almost unchanged. A possible reason for such a selective effect could be the difference in coordination numbers. In the $Pmna$ and $Pmmn$ structures, iron has the same coordination number 6, whereas in the CsCl-type structure it is 8-coordinate. It seems plausible that phase transitions involving coordination number changes are particularly sensitive to electron correlation effects.

Unlike other known FeS phases which are well modeled in the DFT and DFT+U approximations, the CsCl-type structure could only be found to be stable after a thorough investigation with the thermal and electron correlation effects taken into consideration. However, modern computational techniques are capable of dealing with cases as subtle as this, displaying good agreement with experimental results.  

By this means, we gain insight into the intriguing behavior of the iron sulfide exposed to high pressure. The calculated phase diagram gives a clue as to how to synthesize the $Pmmn$ phase which has been predicted recently but still not discovered. Our results show that electron correlations can play an important role even at very high pressures, such as pressures in the Earth’s core, where local magnetic moments on iron atoms are suppressed but magnetic fluctuations are still significant.

The DFT+DMFT calculations are supported by the Russian Science Foundation (grant 14-22-00004). The phonons calculations were performed with the support of the Russian Science Foundation (grant 16-13-10459). The phase diagram was computed with the support of National Science Foundation (grant EAR-1723160).
\bibliography{ref.bib}

\begin{thebibliography}{51}%
\makeatletter
\providecommand \@ifxundefined [1]{%
 \@ifx{#1\undefined}
}%
\providecommand \@ifnum [1]{%
 \ifnum #1\expandafter \@firstoftwo
 \else \expandafter \@secondoftwo
 \fi
}%
\providecommand \@ifx [1]{%
 \ifx #1\expandafter \@firstoftwo
 \else \expandafter \@secondoftwo
 \fi
}%
\providecommand \natexlab [1]{#1}%
\providecommand \enquote  [1]{``#1''}%
\providecommand \bibnamefont  [1]{#1}%
\providecommand \bibfnamefont [1]{#1}%
\providecommand \citenamefont [1]{#1}%
\providecommand \href@noop [0]{\@secondoftwo}%
\providecommand \href [0]{\begingroup \@sanitize@url \@href}%
\providecommand \@href[1]{\@@startlink{#1}\@@href}%
\providecommand \@@href[1]{\endgroup#1\@@endlink}%
\providecommand \@sanitize@url [0]{\catcode `\\12\catcode `\$12\catcode
  `\&12\catcode `\#12\catcode `\^12\catcode `\_12\catcode `\%12\relax}%
\providecommand \@@startlink[1]{}%
\providecommand \@@endlink[0]{}%
\providecommand \url  [0]{\begingroup\@sanitize@url \@url }%
\providecommand \@url [1]{\endgroup\@href {#1}{\urlprefix }}%
\providecommand \urlprefix  [0]{URL }%
\providecommand \Eprint [0]{\href }%
\providecommand \doibase [0]{http://dx.doi.org/}%
\providecommand \selectlanguage [0]{\@gobble}%
\providecommand \bibinfo  [0]{\@secondoftwo}%
\providecommand \bibfield  [0]{\@secondoftwo}%
\providecommand \translation [1]{[#1]}%
\providecommand \BibitemOpen [0]{}%
\providecommand \bibitemStop [0]{}%
\providecommand \bibitemNoStop [0]{.\EOS\space}%
\providecommand \EOS [0]{\spacefactor3000\relax}%
\providecommand \BibitemShut  [1]{\csname bibitem#1\endcsname}%
\let\auto@bib@innerbib\@empty
\bibitem [{\citenamefont {Poirier}(1994)}]{Poirier1994}%
  \BibitemOpen
  \bibfield  {author} {\bibinfo {author} {\bibfnamefont {J.-P.}\ \bibnamefont
  {Poirier}},\ }\href {\doibase 10.1016/0031-9201(94)90120-1} {\bibfield
  {journal} {\bibinfo  {journal} {Physics of the Earth and Planetary
  Interiors}\ }\textbf {\bibinfo {volume} {85}},\ \bibinfo {pages} {319}
  (\bibinfo {year} {1994})}\BibitemShut {NoStop}%
\bibitem [{\citenamefont {Seagle}\ \emph {et~al.}(2006)\citenamefont {Seagle},
  \citenamefont {Campbell}, \citenamefont {Heinz}, \citenamefont {Shen},\ and\
  \citenamefont {Prakapenka}}]{Seagle2006}%
  \BibitemOpen
  \bibfield  {author} {\bibinfo {author} {\bibfnamefont {C.~T.}\ \bibnamefont
  {Seagle}}, \bibinfo {author} {\bibfnamefont {A.~J.}\ \bibnamefont
  {Campbell}}, \bibinfo {author} {\bibfnamefont {D.~L.}\ \bibnamefont {Heinz}},
  \bibinfo {author} {\bibfnamefont {G.}~\bibnamefont {Shen}}, \ and\ \bibinfo
  {author} {\bibfnamefont {V.~B.}\ \bibnamefont {Prakapenka}},\ }\href
  {\doibase 10.1029/2005JB004091} {\bibfield  {journal} {\bibinfo  {journal}
  {Journal of Geophysical Research: Solid Earth}\ }\textbf {\bibinfo {volume}
  {111}},\ \bibinfo {pages} {n/a} (\bibinfo {year} {2006})}\BibitemShut
  {NoStop}%
\bibitem [{\citenamefont {Fei}\ \emph {et~al.}(1995)\citenamefont {Fei},
  \citenamefont {Prewitt}, \citenamefont {Mao},\ and\ \citenamefont
  {Bertka}}]{Fei1995}%
  \BibitemOpen
  \bibfield  {author} {\bibinfo {author} {\bibfnamefont {Y.}~\bibnamefont
  {Fei}}, \bibinfo {author} {\bibfnamefont {C.~T.}\ \bibnamefont {Prewitt}},
  \bibinfo {author} {\bibfnamefont {H.}~\bibnamefont {Mao}}, \ and\ \bibinfo
  {author} {\bibfnamefont {C.~M.}\ \bibnamefont {Bertka}},\ }\href {\doibase
  10.1126/science.268.5219.1892} {\bibfield  {journal} {\bibinfo  {journal}
  {Science}\ }\textbf {\bibinfo {volume} {268}},\ \bibinfo {pages} {1892}
  (\bibinfo {year} {1995})}\BibitemShut {NoStop}%
\bibitem [{\citenamefont {Kamimura}\ \emph {et~al.}(1992)\citenamefont
  {Kamimura}, \citenamefont {Sato}, \citenamefont {Takahashi}, \citenamefont
  {Mori}, \citenamefont {Yoshida},\ and\ \citenamefont
  {Kaneko}}]{Kamimura1992}%
  \BibitemOpen
  \bibfield  {author} {\bibinfo {author} {\bibfnamefont {T.}~\bibnamefont
  {Kamimura}}, \bibinfo {author} {\bibfnamefont {M.}~\bibnamefont {Sato}},
  \bibinfo {author} {\bibfnamefont {H.}~\bibnamefont {Takahashi}}, \bibinfo
  {author} {\bibfnamefont {N.}~\bibnamefont {Mori}}, \bibinfo {author}
  {\bibfnamefont {H.}~\bibnamefont {Yoshida}}, \ and\ \bibinfo {author}
  {\bibfnamefont {T.}~\bibnamefont {Kaneko}},\ }\href {\doibase
  10.1016/0304-8853(92)90787-O} {\bibfield  {journal} {\bibinfo  {journal}
  {Journal of Magnetism and Magnetic Materials}\ }\textbf {\bibinfo {volume}
  {104-107}},\ \bibinfo {pages} {255} (\bibinfo {year} {1992})}\BibitemShut
  {NoStop}%
\bibitem [{\citenamefont {Kusaba}\ \emph {et~al.}(1997)\citenamefont {Kusaba},
  \citenamefont {Syono}, \citenamefont {Kikegawa},\ and\ \citenamefont
  {Simomura}}]{Kusaba1997}%
  \BibitemOpen
  \bibfield  {author} {\bibinfo {author} {\bibfnamefont {K.}~\bibnamefont
  {Kusaba}}, \bibinfo {author} {\bibfnamefont {Y.}~\bibnamefont {Syono}},
  \bibinfo {author} {\bibfnamefont {T.}~\bibnamefont {Kikegawa}}, \ and\
  \bibinfo {author} {\bibfnamefont {O.}~\bibnamefont {Simomura}},\ }\href
  {\doibase 10.1016/S0022-3697(96)00120-5} {\bibfield  {journal} {\bibinfo
  {journal} {Journal of Physics and Chemistry of Solids}\ }\textbf {\bibinfo
  {volume} {58}},\ \bibinfo {pages} {241} (\bibinfo {year} {1997})}\BibitemShut
  {NoStop}%
\bibitem [{\citenamefont {Taylor}\ and\ \citenamefont
  {Mao}(1970)}]{Taylor1970}%
  \BibitemOpen
  \bibfield  {author} {\bibinfo {author} {\bibfnamefont {L.~A.}\ \bibnamefont
  {Taylor}}\ and\ \bibinfo {author} {\bibfnamefont {H.~K.}\ \bibnamefont
  {Mao}},\ }\href@noop {} {\bibfield  {journal} {\bibinfo  {journal} {Science}\
  }\textbf {\bibinfo {volume} {170}},\ \bibinfo {pages} {850} (\bibinfo {year}
  {1970})}\BibitemShut {NoStop}%
\bibitem [{\citenamefont {Wang}\ and\ \citenamefont
  {Salveson}(2005)}]{Wang2005}%
  \BibitemOpen
  \bibfield  {author} {\bibinfo {author} {\bibfnamefont {H.}~\bibnamefont
  {Wang}}\ and\ \bibinfo {author} {\bibfnamefont {I.}~\bibnamefont
  {Salveson}},\ }\href {\doibase 10.1080/01411590500185542} {\bibfield
  {journal} {\bibinfo  {journal} {Phase Transitions}\ }\textbf {\bibinfo
  {volume} {78}},\ \bibinfo {pages} {547} (\bibinfo {year} {2005})}\BibitemShut
  {NoStop}%
\bibitem [{\citenamefont {King}\ and\ \citenamefont
  {Prewitt}(1982)}]{King1982}%
  \BibitemOpen
  \bibfield  {author} {\bibinfo {author} {\bibfnamefont {H.~E.}\ \bibnamefont
  {King}}\ and\ \bibinfo {author} {\bibfnamefont {C.~T.}\ \bibnamefont
  {Prewitt}},\ }\href {\doibase 10.1107/S0567740882007523} {\bibfield
  {journal} {\bibinfo  {journal} {Acta Crystallographica}\ }\textbf {\bibinfo
  {volume} {B38}},\ \bibinfo {pages} {1877} (\bibinfo {year}
  {1982})}\BibitemShut {NoStop}%
\bibitem [{\citenamefont {Sherman}(1995)}]{Sherman1995}%
  \BibitemOpen
  \bibfield  {author} {\bibinfo {author} {\bibfnamefont {D.~M.}\ \bibnamefont
  {Sherman}},\ }\href {\doibase 10.1016/0012-821X(95)00057-J} {\bibfield
  {journal} {\bibinfo  {journal} {Earth and Planetary Science Letters}\
  }\textbf {\bibinfo {volume} {132}},\ \bibinfo {pages} {87} (\bibinfo {year}
  {1995})}\BibitemShut {NoStop}%
\bibitem [{\citenamefont {Kobayashi}\ \emph {et~al.}(1997)\citenamefont
  {Kobayashi}, \citenamefont {Sato}, \citenamefont {Kamimura}, \citenamefont
  {Sakai}, \citenamefont {Onodera}, \citenamefont {Kuroda},\ and\ \citenamefont
  {Yamaguchi}}]{Kobayashi1997}%
  \BibitemOpen
  \bibfield  {author} {\bibinfo {author} {\bibfnamefont {H.}~\bibnamefont
  {Kobayashi}}, \bibinfo {author} {\bibfnamefont {M.}~\bibnamefont {Sato}},
  \bibinfo {author} {\bibfnamefont {T.}~\bibnamefont {Kamimura}}, \bibinfo
  {author} {\bibfnamefont {M.}~\bibnamefont {Sakai}}, \bibinfo {author}
  {\bibfnamefont {H.}~\bibnamefont {Onodera}}, \bibinfo {author} {\bibfnamefont
  {N.}~\bibnamefont {Kuroda}}, \ and\ \bibinfo {author} {\bibfnamefont
  {Y.}~\bibnamefont {Yamaguchi}},\ }\href {\doibase 10.1088/0953-8984/9/2/019}
  {\bibfield  {journal} {\bibinfo  {journal} {Journal of Physics: Condensed
  Matter}\ }\textbf {\bibinfo {volume} {9}},\ \bibinfo {pages} {515} (\bibinfo
  {year} {1997})}\BibitemShut {NoStop}%
\bibitem [{\citenamefont {Alfe}\ and\ \citenamefont {Gillan}(1998)}]{Alfe1998}%
  \BibitemOpen
  \bibfield  {author} {\bibinfo {author} {\bibfnamefont {D.}~\bibnamefont
  {Alfe}}\ and\ \bibinfo {author} {\bibfnamefont {M.~J.}\ \bibnamefont
  {Gillan}},\ }\href {\doibase 10.1103/PhysRevB.58.8248} {\bibfield  {journal}
  {\bibinfo  {journal} {Physical Review B}\ }\textbf {\bibinfo {volume} {58}},\
  \bibinfo {pages} {8248} (\bibinfo {year} {1998})}\BibitemShut {NoStop}%
\bibitem [{\citenamefont {Kusaba}\ \emph {et~al.}(1998)\citenamefont {Kusaba},
  \citenamefont {Syono}, \citenamefont {Kikegawa},\ and\ \citenamefont
  {Shimomura}}]{Kusaba1998}%
  \BibitemOpen
  \bibfield  {author} {\bibinfo {author} {\bibfnamefont {K.}~\bibnamefont
  {Kusaba}}, \bibinfo {author} {\bibfnamefont {Y.}~\bibnamefont {Syono}},
  \bibinfo {author} {\bibfnamefont {T.}~\bibnamefont {Kikegawa}}, \ and\
  \bibinfo {author} {\bibfnamefont {O.}~\bibnamefont {Shimomura}},\ }\href
  {\doibase 10.1016/S0022-3697(98)00015-8} {\bibfield  {journal} {\bibinfo
  {journal} {Journal of Physics and Chemistry of Solids}\ }\textbf {\bibinfo
  {volume} {59}},\ \bibinfo {pages} {945} (\bibinfo {year} {1998})}\BibitemShut
  {NoStop}%
\bibitem [{\citenamefont {Nelmes}\ \emph {et~al.}(1999)\citenamefont {Nelmes},
  \citenamefont {McMahon}, \citenamefont {Belmonte},\ and\ \citenamefont
  {Parise}}]{Nelmes1999}%
  \BibitemOpen
  \bibfield  {author} {\bibinfo {author} {\bibfnamefont {R.~J.}\ \bibnamefont
  {Nelmes}}, \bibinfo {author} {\bibfnamefont {M.~I.}\ \bibnamefont {McMahon}},
  \bibinfo {author} {\bibfnamefont {S.~A.}\ \bibnamefont {Belmonte}}, \ and\
  \bibinfo {author} {\bibfnamefont {J.~B.}\ \bibnamefont {Parise}},\ }\href
  {\doibase 10.1103/PhysRevB.59.9048} {\bibfield  {journal} {\bibinfo
  {journal} {Physical Review B}\ }\textbf {\bibinfo {volume} {59}},\ \bibinfo
  {pages} {9048} (\bibinfo {year} {1999})}\BibitemShut {NoStop}%
\bibitem [{\citenamefont {Takele}\ and\ \citenamefont
  {Hearne}(1999)}]{Takele1999}%
  \BibitemOpen
  \bibfield  {author} {\bibinfo {author} {\bibfnamefont {S.}~\bibnamefont
  {Takele}}\ and\ \bibinfo {author} {\bibfnamefont {G.~R.}\ \bibnamefont
  {Hearne}},\ }\href {\doibase 10.1103/PhysRevB.60.4401} {\bibfield  {journal}
  {\bibinfo  {journal} {Physical Review B}\ }\textbf {\bibinfo {volume} {60}},\
  \bibinfo {pages} {4401} (\bibinfo {year} {1999})}\BibitemShut {NoStop}%
\bibitem [{\citenamefont {Rueff}\ \emph {et~al.}(1999)\citenamefont {Rueff},
  \citenamefont {Kao}, \citenamefont {Struzhkin}, \citenamefont {Badro},
  \citenamefont {Shu}, \citenamefont {Hemley},\ and\ \citenamefont
  {Mao}}]{Rueff1999}%
  \BibitemOpen
  \bibfield  {author} {\bibinfo {author} {\bibfnamefont {J.-P.}\ \bibnamefont
  {Rueff}}, \bibinfo {author} {\bibfnamefont {C.-C.}\ \bibnamefont {Kao}},
  \bibinfo {author} {\bibfnamefont {V.~V.}\ \bibnamefont {Struzhkin}}, \bibinfo
  {author} {\bibfnamefont {J.}~\bibnamefont {Badro}}, \bibinfo {author}
  {\bibfnamefont {J.}~\bibnamefont {Shu}}, \bibinfo {author} {\bibfnamefont
  {R.~J.}\ \bibnamefont {Hemley}}, \ and\ \bibinfo {author} {\bibfnamefont
  {H.~K.}\ \bibnamefont {Mao}},\ }\href {\doibase 10.1103/PhysRevLett.82.3284}
  {\bibfield  {journal} {\bibinfo  {journal} {Physical Review Letters}\
  }\textbf {\bibinfo {volume} {82}},\ \bibinfo {pages} {3284} (\bibinfo {year}
  {1999})}\BibitemShut {NoStop}%
\bibitem [{\citenamefont {Vo{\v{c}}adlo}\ \emph {et~al.}(2000)\citenamefont
  {Vo{\v{c}}adlo}, \citenamefont {Alf{\`{e}}}, \citenamefont {Price},\ and\
  \citenamefont {Gillan}}]{Vocadlo2000}%
  \BibitemOpen
  \bibfield  {author} {\bibinfo {author} {\bibfnamefont {L.}~\bibnamefont
  {Vo{\v{c}}adlo}}, \bibinfo {author} {\bibfnamefont {D.}~\bibnamefont
  {Alf{\`{e}}}}, \bibinfo {author} {\bibfnamefont {G.~D.}\ \bibnamefont
  {Price}}, \ and\ \bibinfo {author} {\bibfnamefont {M.~J.}\ \bibnamefont
  {Gillan}},\ }\href {\doibase 10.1016/S0031-9201(00)00151-5} {\bibfield
  {journal} {\bibinfo  {journal} {Physics of the Earth and Planetary
  Interiors}\ }\textbf {\bibinfo {volume} {120}},\ \bibinfo {pages} {145}
  (\bibinfo {year} {2000})}\BibitemShut {NoStop}%
\bibitem [{\citenamefont {Marshall}\ \emph {et~al.}(2000)\citenamefont
  {Marshall}, \citenamefont {Nelmes}, \citenamefont {Loveday}, \citenamefont
  {Klotz}, \citenamefont {Besson}, \citenamefont {Hamel},\ and\ \citenamefont
  {Parise}}]{Marshall2000}%
  \BibitemOpen
  \bibfield  {author} {\bibinfo {author} {\bibfnamefont {W.~G.}\ \bibnamefont
  {Marshall}}, \bibinfo {author} {\bibfnamefont {R.~J.}\ \bibnamefont
  {Nelmes}}, \bibinfo {author} {\bibfnamefont {J.~S.}\ \bibnamefont {Loveday}},
  \bibinfo {author} {\bibfnamefont {S.}~\bibnamefont {Klotz}}, \bibinfo
  {author} {\bibfnamefont {J.~M.}\ \bibnamefont {Besson}}, \bibinfo {author}
  {\bibfnamefont {G.}~\bibnamefont {Hamel}}, \ and\ \bibinfo {author}
  {\bibfnamefont {J.~B.}\ \bibnamefont {Parise}},\ }\href {\doibase
  10.1103/PhysRevB.61.11201} {\bibfield  {journal} {\bibinfo  {journal}
  {Physical Review B}\ }\textbf {\bibinfo {volume} {61}},\ \bibinfo {pages}
  {11201} (\bibinfo {year} {2000})}\BibitemShut {NoStop}%
\bibitem [{\citenamefont {Kavner}\ \emph {et~al.}(2000)\citenamefont {Kavner},
  \citenamefont {Duffy},\ and\ \citenamefont {Shen}}]{Kavner2000}%
  \BibitemOpen
  \bibfield  {author} {\bibinfo {author} {\bibfnamefont {A.}~\bibnamefont
  {Kavner}}, \bibinfo {author} {\bibfnamefont {T.~S.}\ \bibnamefont {Duffy}}, \
  and\ \bibinfo {author} {\bibfnamefont {G.}~\bibnamefont {Shen}},\ }\href
  {\doibase 10.1016/S0012-821X(00)00356-3} {\bibfield  {journal} {\bibinfo
  {journal} {Earth and Planetary Science Letters}\ }\textbf {\bibinfo {volume}
  {5707}},\ \bibinfo {pages} {1} (\bibinfo {year} {2000})}\BibitemShut
  {NoStop}%
\bibitem [{\citenamefont {Martin}\ \emph {et~al.}(2001)\citenamefont {Martin},
  \citenamefont {Price},\ and\ \citenamefont {Vo{\v{c}}adlo}}]{Martin2001}%
  \BibitemOpen
  \bibfield  {author} {\bibinfo {author} {\bibfnamefont {P.}~\bibnamefont
  {Martin}}, \bibinfo {author} {\bibfnamefont {G.}~\bibnamefont {Price}}, \
  and\ \bibinfo {author} {\bibfnamefont {L.}~\bibnamefont {Vo{\v{c}}adlo}},\
  }\href {\doibase 10.1180/002646101550217} {\bibfield  {journal} {\bibinfo
  {journal} {Mineralogical Magazine}\ }\textbf {\bibinfo {volume} {65}},\
  \bibinfo {pages} {181} (\bibinfo {year} {2001})}\BibitemShut {NoStop}%
\bibitem [{\citenamefont {Urakawa}\ \emph {et~al.}(2004)\citenamefont
  {Urakawa}, \citenamefont {Someya}, \citenamefont {Terasaki}, \citenamefont
  {Katsura}, \citenamefont {Yokoshi}, \citenamefont {ichi Funakoshi},
  \citenamefont {Utsumi}, \citenamefont {Katayama}, \citenamefont {ichiro
  Sueda},\ and\ \citenamefont {Irifune}}]{Urakawa2004}%
  \BibitemOpen
  \bibfield  {author} {\bibinfo {author} {\bibfnamefont {S.}~\bibnamefont
  {Urakawa}}, \bibinfo {author} {\bibfnamefont {K.}~\bibnamefont {Someya}},
  \bibinfo {author} {\bibfnamefont {H.}~\bibnamefont {Terasaki}}, \bibinfo
  {author} {\bibfnamefont {T.}~\bibnamefont {Katsura}}, \bibinfo {author}
  {\bibfnamefont {S.}~\bibnamefont {Yokoshi}}, \bibinfo {author} {\bibfnamefont
  {K.}~\bibnamefont {ichi Funakoshi}}, \bibinfo {author} {\bibfnamefont
  {W.}~\bibnamefont {Utsumi}}, \bibinfo {author} {\bibfnamefont
  {Y.}~\bibnamefont {Katayama}}, \bibinfo {author} {\bibfnamefont
  {Y.}~\bibnamefont {ichiro Sueda}}, \ and\ \bibinfo {author} {\bibfnamefont
  {T.}~\bibnamefont {Irifune}},\ }\href {\doibase 10.1016/j.pepi.2003.12.015}
  {\bibfield  {journal} {\bibinfo  {journal} {Physics of the Earth and
  Planetary Interiors}\ }\textbf {\bibinfo {volume} {143-144}},\ \bibinfo
  {pages} {469} (\bibinfo {year} {2004})}\BibitemShut {NoStop}%
\bibitem [{\citenamefont {Kobayashi}\ \emph {et~al.}(2004)\citenamefont
  {Kobayashi}, \citenamefont {Kamimura}, \citenamefont {Alfe}, \citenamefont
  {Sturhahn}, \citenamefont {Zhao},\ and\ \citenamefont {Alp}}]{Kobayashi2004}%
  \BibitemOpen
  \bibfield  {author} {\bibinfo {author} {\bibfnamefont {H.}~\bibnamefont
  {Kobayashi}}, \bibinfo {author} {\bibfnamefont {T.}~\bibnamefont {Kamimura}},
  \bibinfo {author} {\bibfnamefont {D.}~\bibnamefont {Alfe}}, \bibinfo {author}
  {\bibfnamefont {W.}~\bibnamefont {Sturhahn}}, \bibinfo {author}
  {\bibfnamefont {J.}~\bibnamefont {Zhao}}, \ and\ \bibinfo {author}
  {\bibfnamefont {E.~E.}\ \bibnamefont {Alp}},\ }\href {\doibase
  10.1103/PhysRevLett.93.195503} {\bibfield  {journal} {\bibinfo  {journal}
  {Physical Review Letters}\ }\textbf {\bibinfo {volume} {93}},\ \bibinfo
  {pages} {195503} (\bibinfo {year} {2004})}\BibitemShut {NoStop}%
\bibitem [{\citenamefont {Ono}\ and\ \citenamefont {Kikegawa}(2006)}]{Ono2006}%
  \BibitemOpen
  \bibfield  {author} {\bibinfo {author} {\bibfnamefont {S.}~\bibnamefont
  {Ono}}\ and\ \bibinfo {author} {\bibfnamefont {T.}~\bibnamefont {Kikegawa}},\
  }\href {\doibase 10.2138/am.2006.2347} {\bibfield  {journal} {\bibinfo
  {journal} {American Mineralogist}\ }\textbf {\bibinfo {volume} {91}},\
  \bibinfo {pages} {1941} (\bibinfo {year} {2006})}\BibitemShut {NoStop}%
\bibitem [{\citenamefont {Ushakov}\ \emph {et~al.}(2017)\citenamefont
  {Ushakov}, \citenamefont {Shorikov}, \citenamefont {Anisimov}, \citenamefont
  {Baranov},\ and\ \citenamefont {Streltsov}}]{Ushakov2017}%
  \BibitemOpen
  \bibfield  {author} {\bibinfo {author} {\bibfnamefont {A.~V.}\ \bibnamefont
  {Ushakov}}, \bibinfo {author} {\bibfnamefont {A.~O.}\ \bibnamefont
  {Shorikov}}, \bibinfo {author} {\bibfnamefont {V.~I.}\ \bibnamefont
  {Anisimov}}, \bibinfo {author} {\bibfnamefont {N.~V.}\ \bibnamefont
  {Baranov}}, \ and\ \bibinfo {author} {\bibfnamefont {S.~V.}\ \bibnamefont
  {Streltsov}},\ }\href {\doibase 10.1103/PhysRevB.95.205116} {\bibfield
  {journal} {\bibinfo  {journal} {Physical Review B}\ }\textbf {\bibinfo
  {volume} {95}},\ \bibinfo {pages} {205116} (\bibinfo {year}
  {2017})}\BibitemShut {NoStop}%
\bibitem [{\citenamefont {Keller-Besrest}\ and\ \citenamefont
  {Collin}(1990)}]{Keller1990}%
  \BibitemOpen
  \bibfield  {author} {\bibinfo {author} {\bibfnamefont {F.}~\bibnamefont
  {Keller-Besrest}}\ and\ \bibinfo {author} {\bibfnamefont {G.}~\bibnamefont
  {Collin}},\ }\href {\doibase 10.1016/0022-4596(90)90320-W} {\bibfield
  {journal} {\bibinfo  {journal} {Journal of Solid State Chemistry}\ }\textbf
  {\bibinfo {volume} {84}},\ \bibinfo {pages} {211} (\bibinfo {year}
  {1990})}\BibitemShut {NoStop}%
\bibitem [{\citenamefont {King}\ \emph {et~al.}(1978)\citenamefont {King},
  \citenamefont {Virgo},\ and\ \citenamefont {Mao}}]{King1978}%
  \BibitemOpen
  \bibfield  {author} {\bibinfo {author} {\bibfnamefont {H.~E.}\ \bibnamefont
  {King}}, \bibinfo {author} {\bibfnamefont {D.}~\bibnamefont {Virgo}}, \ and\
  \bibinfo {author} {\bibfnamefont {H.~K.}\ \bibnamefont {Mao}},\ }\href@noop
  {} {\bibfield  {journal} {\bibinfo  {journal} {Carnegie Inst. Wash. Year
  Book}\ }\textbf {\bibinfo {volume} {58}},\ \bibinfo {pages} {241} (\bibinfo
  {year} {1978})}\BibitemShut {NoStop}%
\bibitem [{\citenamefont {Ono}\ \emph {et~al.}(2007)\citenamefont {Ono},
  \citenamefont {Kikegawa},\ and\ \citenamefont {Ohishi}}]{Ono2007}%
  \BibitemOpen
  \bibfield  {author} {\bibinfo {author} {\bibfnamefont {S.}~\bibnamefont
  {Ono}}, \bibinfo {author} {\bibfnamefont {T.}~\bibnamefont {Kikegawa}}, \
  and\ \bibinfo {author} {\bibfnamefont {Y.}~\bibnamefont {Ohishi}},\ }\href
  {\doibase 10.1127/0935-1221/2007/0019-1713} {\bibfield  {journal} {\bibinfo
  {journal} {European Journal of Mineralogy}\ }\textbf {\bibinfo {volume}
  {19}},\ \bibinfo {pages} {183} (\bibinfo {year} {2007})}\BibitemShut
  {NoStop}%
\bibitem [{\citenamefont {Sata}\ \emph {et~al.}(2008)\citenamefont {Sata},
  \citenamefont {Ohfuji}, \citenamefont {Hirose}, \citenamefont {Kobayashi},
  \citenamefont {Ohishi},\ and\ \citenamefont {Hirao}}]{Sata2008}%
  \BibitemOpen
  \bibfield  {author} {\bibinfo {author} {\bibfnamefont {N.}~\bibnamefont
  {Sata}}, \bibinfo {author} {\bibfnamefont {H.}~\bibnamefont {Ohfuji}},
  \bibinfo {author} {\bibfnamefont {K.}~\bibnamefont {Hirose}}, \bibinfo
  {author} {\bibfnamefont {H.}~\bibnamefont {Kobayashi}}, \bibinfo {author}
  {\bibfnamefont {Y.}~\bibnamefont {Ohishi}}, \ and\ \bibinfo {author}
  {\bibfnamefont {N.}~\bibnamefont {Hirao}},\ }\href {\doibase
  10.2138/am.2008.2762} {\bibfield  {journal} {\bibinfo  {journal} {American
  Mineralogist}\ }\textbf {\bibinfo {volume} {93}},\ \bibinfo {pages} {492}
  (\bibinfo {year} {2008})}\BibitemShut {NoStop}%
\bibitem [{Note1()}]{Note1}%
  \BibitemOpen
  \bibinfo {note} {The CsCl-phase was synthesized at 186 GPa after heating to
  1300 K, and this P,T-point (shown by red star on our phase diagram)
  corresponds to stability of this phase. Then this phase was quenched to 298
  K, where it probably is metastable.}\BibitemShut {Stop}%
\bibitem [{\citenamefont {Ono}\ \emph {et~al.}(2008)\citenamefont {Ono},
  \citenamefont {Oganov}, \citenamefont {Brodholt}, \citenamefont
  {Vo{\v{c}}adlo}, \citenamefont {Wood}, \citenamefont {Lyakhov}, \citenamefont
  {Glass}, \citenamefont {C{\^{o}}t{\'{e}}},\ and\ \citenamefont
  {Price}}]{Ono2008}%
  \BibitemOpen
  \bibfield  {author} {\bibinfo {author} {\bibfnamefont {S.}~\bibnamefont
  {Ono}}, \bibinfo {author} {\bibfnamefont {A.~R.}\ \bibnamefont {Oganov}},
  \bibinfo {author} {\bibfnamefont {J.~P.}\ \bibnamefont {Brodholt}}, \bibinfo
  {author} {\bibfnamefont {L.}~\bibnamefont {Vo{\v{c}}adlo}}, \bibinfo {author}
  {\bibfnamefont {I.~G.}\ \bibnamefont {Wood}}, \bibinfo {author}
  {\bibfnamefont {A.}~\bibnamefont {Lyakhov}}, \bibinfo {author} {\bibfnamefont
  {C.~W.}\ \bibnamefont {Glass}}, \bibinfo {author} {\bibfnamefont {A.~S.}\
  \bibnamefont {C{\^{o}}t{\'{e}}}}, \ and\ \bibinfo {author} {\bibfnamefont
  {G.~D.}\ \bibnamefont {Price}},\ }\href {\doibase 10.1016/j.epsl.2008.05.017}
  {\bibfield  {journal} {\bibinfo  {journal} {Earth and Planetary Science
  Letters}\ }\textbf {\bibinfo {volume} {272}},\ \bibinfo {pages} {481}
  (\bibinfo {year} {2008})}\BibitemShut {NoStop}%
\bibitem [{\citenamefont {Dobson}\ \emph {et~al.}(2002)\citenamefont {Dobson},
  \citenamefont {Vo{\v{c}}adlo},\ and\ \citenamefont {Wood}}]{Dobson2002}%
  \BibitemOpen
  \bibfield  {author} {\bibinfo {author} {\bibfnamefont {D.~P.}\ \bibnamefont
  {Dobson}}, \bibinfo {author} {\bibfnamefont {L.}~\bibnamefont
  {Vo{\v{c}}adlo}}, \ and\ \bibinfo {author} {\bibfnamefont {I.~G.}\
  \bibnamefont {Wood}},\ }\href {\doibase 10.2138/am-2002-5-623} {\bibfield
  {journal} {\bibinfo  {journal} {American Mineralogist}\ }\textbf {\bibinfo
  {volume} {87}},\ \bibinfo {pages} {784} (\bibinfo {year} {2002})}\BibitemShut
  {NoStop}%
\bibitem [{\citenamefont {Bazhanova}\ \emph {et~al.}(2017)\citenamefont
  {Bazhanova}, \citenamefont {Roizen},\ and\ \citenamefont
  {Oganov}}]{Bazhanova2017}%
  \BibitemOpen
  \bibfield  {author} {\bibinfo {author} {\bibfnamefont {Z.}~\bibnamefont
  {Bazhanova}}, \bibinfo {author} {\bibfnamefont {V.}~\bibnamefont {Roizen}}, \
  and\ \bibinfo {author} {\bibfnamefont {A.}~\bibnamefont {Oganov}},\ }\href
  {\doibase 10.3367/UFNe.2017.03.038079} {\bibfield  {journal} {\bibinfo
  {journal} {Uspekhi Fizicheskih Nauk}\ } (\bibinfo {year} {2017}),\
  10.3367/UFNe.2017.03.038079}\BibitemShut {NoStop}%
\bibitem [{\citenamefont {Momma}\ and\ \citenamefont
  {Izumi}(2011)}]{Momma2011}%
  \BibitemOpen
  \bibfield  {author} {\bibinfo {author} {\bibfnamefont {K.}~\bibnamefont
  {Momma}}\ and\ \bibinfo {author} {\bibfnamefont {F.}~\bibnamefont {Izumi}},\
  }\href {\doibase https://doi.org/10.1107/S0021889811038970} {\bibfield
  {journal} {\bibinfo  {journal} {J. Appl. Cryst.}\ }\textbf {\bibinfo {volume}
  {44}},\ \bibinfo {pages} {1272} (\bibinfo {year} {2011})}\BibitemShut
  {NoStop}%
\bibitem [{\citenamefont {Kresse}\ and\ \citenamefont
  {Furthm{\"{u}}ller}(1996)}]{Kresse1996}%
  \BibitemOpen
  \bibfield  {author} {\bibinfo {author} {\bibfnamefont {G.}~\bibnamefont
  {Kresse}}\ and\ \bibinfo {author} {\bibfnamefont {J.}~\bibnamefont
  {Furthm{\"{u}}ller}},\ }\href {\doibase 10.1103/PhysRevB.54.11169} {\bibfield
   {journal} {\bibinfo  {journal} {Physical Review B}\ }\textbf {\bibinfo
  {volume} {54}},\ \bibinfo {pages} {11169} (\bibinfo {year}
  {1996})}\BibitemShut {NoStop}%
\bibitem [{\citenamefont {Perdew}\ \emph {et~al.}(1996)\citenamefont {Perdew},
  \citenamefont {Burke},\ and\ \citenamefont {Ernzerhof}}]{Perdew1996}%
  \BibitemOpen
  \bibfield  {author} {\bibinfo {author} {\bibfnamefont {J.}~\bibnamefont
  {Perdew}}, \bibinfo {author} {\bibfnamefont {K.}~\bibnamefont {Burke}}, \
  and\ \bibinfo {author} {\bibfnamefont {M.}~\bibnamefont {Ernzerhof}},\ }\href
  {http://www.ncbi.nlm.nih.gov/pubmed/10062328} {\bibfield  {journal} {\bibinfo
   {journal} {Physical Review Letters}\ }\textbf {\bibinfo {volume} {77}},\
  \bibinfo {pages} {3865} (\bibinfo {year} {1996})}\BibitemShut {NoStop}%
\bibitem [{\citenamefont {Anisimov}\ \emph {et~al.}(1997)\citenamefont
  {Anisimov}, \citenamefont {Poteryaev}, \citenamefont {Korotin}, \citenamefont
  {Anokhin},\ and\ \citenamefont {Kotliar}}]{Anisimov1997}%
  \BibitemOpen
  \bibfield  {author} {\bibinfo {author} {\bibfnamefont {V.~I.}\ \bibnamefont
  {Anisimov}}, \bibinfo {author} {\bibfnamefont {A.~I.}\ \bibnamefont
  {Poteryaev}}, \bibinfo {author} {\bibfnamefont {M.~A.}\ \bibnamefont
  {Korotin}}, \bibinfo {author} {\bibfnamefont {A.~O.}\ \bibnamefont
  {Anokhin}}, \ and\ \bibinfo {author} {\bibfnamefont {G.}~\bibnamefont
  {Kotliar}},\ }\href {\doibase 10.1088/0953-8984/9/35/010} {\bibfield
  {journal} {\bibinfo  {journal} {Journal of Physics: Condensed Matter}\
  }\textbf {\bibinfo {volume} {9}},\ \bibinfo {pages} {7359} (\bibinfo {year}
  {1997})}\BibitemShut {NoStop}%
\bibitem [{\citenamefont {Held}\ \emph {et~al.}(2006)\citenamefont {Held},
  \citenamefont {Nekrasov}, \citenamefont {Keller}, \citenamefont {Eyert},
  \citenamefont {Bl{\"{u}}mer}, \citenamefont {McMahan}, \citenamefont
  {Scalettar}, \citenamefont {Pruschke}, \citenamefont {Anisimov},\ and\
  \citenamefont {Vollhardt}}]{Held2006}%
  \BibitemOpen
  \bibfield  {author} {\bibinfo {author} {\bibfnamefont {K.}~\bibnamefont
  {Held}}, \bibinfo {author} {\bibfnamefont {I.~A.}\ \bibnamefont {Nekrasov}},
  \bibinfo {author} {\bibfnamefont {G.}~\bibnamefont {Keller}}, \bibinfo
  {author} {\bibfnamefont {V.}~\bibnamefont {Eyert}}, \bibinfo {author}
  {\bibfnamefont {N.}~\bibnamefont {Bl{\"{u}}mer}}, \bibinfo {author}
  {\bibfnamefont {A.~K.}\ \bibnamefont {McMahan}}, \bibinfo {author}
  {\bibfnamefont {R.~T.}\ \bibnamefont {Scalettar}}, \bibinfo {author}
  {\bibfnamefont {T.}~\bibnamefont {Pruschke}}, \bibinfo {author}
  {\bibfnamefont {V.~I.}\ \bibnamefont {Anisimov}}, \ and\ \bibinfo {author}
  {\bibfnamefont {D.}~\bibnamefont {Vollhardt}},\ }\href {\doibase
  10.1002/pssb.200642053} {\bibfield  {journal} {\bibinfo  {journal} {physica
  status solidi (b)}\ }\textbf {\bibinfo {volume} {243}},\ \bibinfo {pages}
  {2599} (\bibinfo {year} {2006})}\BibitemShut {NoStop}%
\bibitem [{\citenamefont {Shorikov}\ \emph {et~al.}(2010)\citenamefont
  {Shorikov}, \citenamefont {Pchelkina}, \citenamefont {Anisimov},
  \citenamefont {Skornyakov},\ and\ \citenamefont {Korotin}}]{Shorikov2010a}%
  \BibitemOpen
  \bibfield  {author} {\bibinfo {author} {\bibfnamefont {A.}~\bibnamefont
  {Shorikov}}, \bibinfo {author} {\bibfnamefont {Z.}~\bibnamefont {Pchelkina}},
  \bibinfo {author} {\bibfnamefont {V.}~\bibnamefont {Anisimov}}, \bibinfo
  {author} {\bibfnamefont {S.}~\bibnamefont {Skornyakov}}, \ and\ \bibinfo
  {author} {\bibfnamefont {M.}~\bibnamefont {Korotin}},\ }\href {\doibase
  10.1103/PhysRevB.82.195101} {\bibfield  {journal} {\bibinfo  {journal}
  {Physical Review B}\ }\textbf {\bibinfo {volume} {82}},\ \bibinfo {pages}
  {195101} (\bibinfo {year} {2010})}\BibitemShut {NoStop}%
\bibitem [{\citenamefont {Kunes}\ \emph {et~al.}(2008)\citenamefont {Kunes},
  \citenamefont {Lukoyanov}, \citenamefont {Anisimov}, \citenamefont
  {Scalettar},\ and\ \citenamefont {Pickett}}]{Kunes2008}%
  \BibitemOpen
  \bibfield  {author} {\bibinfo {author} {\bibfnamefont {J.}~\bibnamefont
  {Kunes}}, \bibinfo {author} {\bibfnamefont {A.~V.}\ \bibnamefont
  {Lukoyanov}}, \bibinfo {author} {\bibfnamefont {V.~I.}\ \bibnamefont
  {Anisimov}}, \bibinfo {author} {\bibfnamefont {R.~T.}\ \bibnamefont
  {Scalettar}}, \ and\ \bibinfo {author} {\bibfnamefont {W.~E.}\ \bibnamefont
  {Pickett}},\ }\href {\doibase 10.1038/nmat2115} {\bibfield  {journal}
  {\bibinfo  {journal} {Nature materials}\ }\textbf {\bibinfo {volume} {7}},\
  \bibinfo {pages} {198} (\bibinfo {year} {2008})}\BibitemShut {NoStop}%
\bibitem [{\citenamefont {Shorikov}\ \emph {et~al.}(2015)\citenamefont
  {Shorikov}, \citenamefont {Lukoyanov}, \citenamefont {Anisimov},\ and\
  \citenamefont {Savrasov}}]{Shorikov2015a}%
  \BibitemOpen
  \bibfield  {author} {\bibinfo {author} {\bibfnamefont {A.~O.}\ \bibnamefont
  {Shorikov}}, \bibinfo {author} {\bibfnamefont {A.~V.}\ \bibnamefont
  {Lukoyanov}}, \bibinfo {author} {\bibfnamefont {V.~I.}\ \bibnamefont
  {Anisimov}}, \ and\ \bibinfo {author} {\bibfnamefont {S.~Y.}\ \bibnamefont
  {Savrasov}},\ }\href {\doibase 10.1103/PhysRevB.92.035125} {\bibfield
  {journal} {\bibinfo  {journal} {Physical Review B}\ }\textbf {\bibinfo
  {volume} {92}},\ \bibinfo {pages} {035125} (\bibinfo {year}
  {2015})}\BibitemShut {NoStop}%
\bibitem [{\citenamefont {Skorikov}\ \emph {et~al.}(2015)\citenamefont
  {Skorikov}, \citenamefont {Shorikov}, \citenamefont {Skornyakov},
  \citenamefont {Korotin},\ and\ \citenamefont {Anisimov}}]{Skorikov2015}%
  \BibitemOpen
  \bibfield  {author} {\bibinfo {author} {\bibfnamefont {N.~A.}\ \bibnamefont
  {Skorikov}}, \bibinfo {author} {\bibfnamefont {A.~O.}\ \bibnamefont
  {Shorikov}}, \bibinfo {author} {\bibfnamefont {S.~L.}\ \bibnamefont
  {Skornyakov}}, \bibinfo {author} {\bibfnamefont {M.~A.}\ \bibnamefont
  {Korotin}}, \ and\ \bibinfo {author} {\bibfnamefont {V.~I.}\ \bibnamefont
  {Anisimov}},\ }\href {\doibase 10.1088/0953-8984/27/27/275501} {\bibfield
  {journal} {\bibinfo  {journal} {Journal of Physics: Condensed Matter}\
  }\textbf {\bibinfo {volume} {27}},\ \bibinfo {pages} {275501} (\bibinfo
  {year} {2015})}\BibitemShut {NoStop}%
\bibitem [{\citenamefont {Giannozzi}\ \emph {et~al.}(2009)\citenamefont
  {Giannozzi}, \citenamefont {Baroni}, \citenamefont {Bonini}, \citenamefont
  {Calandra}, \citenamefont {Car}, \citenamefont {Cavazzoni}, \citenamefont
  {Ceresoli}, \citenamefont {Chiarotti}, \citenamefont {Cococcioni},
  \citenamefont {Dabo}, \citenamefont {{Dal Corso}}, \citenamefont
  {de~Gironcoli}, \citenamefont {Fabris}, \citenamefont {Fratesi},
  \citenamefont {Gebauer}, \citenamefont {Gerstmann}, \citenamefont
  {Gougoussis}, \citenamefont {Kokalj}, \citenamefont {Lazzeri}, \citenamefont
  {Martin-Samos}, \citenamefont {Marzari}, \citenamefont {Mauri}, \citenamefont
  {Mazzarello}, \citenamefont {Paolini}, \citenamefont {Pasquarello},
  \citenamefont {Paulatto}, \citenamefont {Sbraccia}, \citenamefont {Scandolo},
  \citenamefont {Sclauzero}, \citenamefont {Seitsonen}, \citenamefont
  {Smogunov}, \citenamefont {Umari},\ and\ \citenamefont
  {Wentzcovitch}}]{Giannozzi2009}%
  \BibitemOpen
  \bibfield  {author} {\bibinfo {author} {\bibfnamefont {P.}~\bibnamefont
  {Giannozzi}}, \bibinfo {author} {\bibfnamefont {S.}~\bibnamefont {Baroni}},
  \bibinfo {author} {\bibfnamefont {N.}~\bibnamefont {Bonini}}, \bibinfo
  {author} {\bibfnamefont {M.}~\bibnamefont {Calandra}}, \bibinfo {author}
  {\bibfnamefont {R.}~\bibnamefont {Car}}, \bibinfo {author} {\bibfnamefont
  {C.}~\bibnamefont {Cavazzoni}}, \bibinfo {author} {\bibfnamefont
  {D.}~\bibnamefont {Ceresoli}}, \bibinfo {author} {\bibfnamefont {G.~L.}\
  \bibnamefont {Chiarotti}}, \bibinfo {author} {\bibfnamefont {M.}~\bibnamefont
  {Cococcioni}}, \bibinfo {author} {\bibfnamefont {I.}~\bibnamefont {Dabo}},
  \bibinfo {author} {\bibfnamefont {A.}~\bibnamefont {{Dal Corso}}}, \bibinfo
  {author} {\bibfnamefont {S.}~\bibnamefont {de~Gironcoli}}, \bibinfo {author}
  {\bibfnamefont {S.}~\bibnamefont {Fabris}}, \bibinfo {author} {\bibfnamefont
  {G.}~\bibnamefont {Fratesi}}, \bibinfo {author} {\bibfnamefont
  {R.}~\bibnamefont {Gebauer}}, \bibinfo {author} {\bibfnamefont
  {U.}~\bibnamefont {Gerstmann}}, \bibinfo {author} {\bibfnamefont
  {C.}~\bibnamefont {Gougoussis}}, \bibinfo {author} {\bibfnamefont
  {A.}~\bibnamefont {Kokalj}}, \bibinfo {author} {\bibfnamefont
  {M.}~\bibnamefont {Lazzeri}}, \bibinfo {author} {\bibfnamefont
  {L.}~\bibnamefont {Martin-Samos}}, \bibinfo {author} {\bibfnamefont
  {N.}~\bibnamefont {Marzari}}, \bibinfo {author} {\bibfnamefont
  {F.}~\bibnamefont {Mauri}}, \bibinfo {author} {\bibfnamefont
  {R.}~\bibnamefont {Mazzarello}}, \bibinfo {author} {\bibfnamefont
  {S.}~\bibnamefont {Paolini}}, \bibinfo {author} {\bibfnamefont
  {A.}~\bibnamefont {Pasquarello}}, \bibinfo {author} {\bibfnamefont
  {L.}~\bibnamefont {Paulatto}}, \bibinfo {author} {\bibfnamefont
  {C.}~\bibnamefont {Sbraccia}}, \bibinfo {author} {\bibfnamefont
  {S.}~\bibnamefont {Scandolo}}, \bibinfo {author} {\bibfnamefont
  {G.}~\bibnamefont {Sclauzero}}, \bibinfo {author} {\bibfnamefont {A.~P.}\
  \bibnamefont {Seitsonen}}, \bibinfo {author} {\bibfnamefont {A.}~\bibnamefont
  {Smogunov}}, \bibinfo {author} {\bibfnamefont {P.}~\bibnamefont {Umari}}, \
  and\ \bibinfo {author} {\bibfnamefont {R.~M.}\ \bibnamefont {Wentzcovitch}},\
  }\href {\doibase 10.1088/0953-8984/21/39/395502} {\bibfield  {journal}
  {\bibinfo  {journal} {Journal of Physics: Condensed Matter}\ }\textbf
  {\bibinfo {volume} {21}},\ \bibinfo {pages} {395502} (\bibinfo {year}
  {2009})}\BibitemShut {NoStop}%
\bibitem [{\citenamefont {Korotin}\ \emph {et~al.}(2008)\citenamefont
  {Korotin}, \citenamefont {Kozhevnikov}, \citenamefont {Skornyakov},
  \citenamefont {Leonov}, \citenamefont {Binggeli}, \citenamefont {Anisimov},\
  and\ \citenamefont {Trimarchi}}]{Korotin2008}%
  \BibitemOpen
  \bibfield  {author} {\bibinfo {author} {\bibfnamefont {D.}~\bibnamefont
  {Korotin}}, \bibinfo {author} {\bibfnamefont {A.~V.}\ \bibnamefont
  {Kozhevnikov}}, \bibinfo {author} {\bibfnamefont {S.~L.}\ \bibnamefont
  {Skornyakov}}, \bibinfo {author} {\bibfnamefont {I.}~\bibnamefont {Leonov}},
  \bibinfo {author} {\bibfnamefont {N.}~\bibnamefont {Binggeli}}, \bibinfo
  {author} {\bibfnamefont {V.~I.}\ \bibnamefont {Anisimov}}, \ and\ \bibinfo
  {author} {\bibfnamefont {G.}~\bibnamefont {Trimarchi}},\ }\href {\doibase
  10.1140/epjb/e2008-00326-3} {\bibfield  {journal} {\bibinfo  {journal} {The
  European Physical Journal B}\ }\textbf {\bibinfo {volume} {65}},\ \bibinfo
  {pages} {91} (\bibinfo {year} {2008})}\BibitemShut {NoStop}%
\bibitem [{\citenamefont {Lichtenstein}\ and\ \citenamefont
  {Katsnelson}(1998)}]{Lichtenstein1998}%
  \BibitemOpen
  \bibfield  {author} {\bibinfo {author} {\bibfnamefont {A.~I.}\ \bibnamefont
  {Lichtenstein}}\ and\ \bibinfo {author} {\bibfnamefont {M.~I.}\ \bibnamefont
  {Katsnelson}},\ }\href {\doibase 10.1103/PhysRevB.57.6884} {\bibfield
  {journal} {\bibinfo  {journal} {Physical Review B}\ }\textbf {\bibinfo
  {volume} {57}},\ \bibinfo {pages} {6884} (\bibinfo {year}
  {1998})}\BibitemShut {NoStop}%
\bibitem [{\citenamefont {Gull}\ \emph {et~al.}(2011)\citenamefont {Gull},
  \citenamefont {Millis}, \citenamefont {Lichtenstein}, \citenamefont
  {Rubtsov}, \citenamefont {Troyer},\ and\ \citenamefont {Werner}}]{Gull2011}%
  \BibitemOpen
  \bibfield  {author} {\bibinfo {author} {\bibfnamefont {E.}~\bibnamefont
  {Gull}}, \bibinfo {author} {\bibfnamefont {A.~J.}\ \bibnamefont {Millis}},
  \bibinfo {author} {\bibfnamefont {A.~I.}\ \bibnamefont {Lichtenstein}},
  \bibinfo {author} {\bibfnamefont {A.~N.}\ \bibnamefont {Rubtsov}}, \bibinfo
  {author} {\bibfnamefont {M.}~\bibnamefont {Troyer}}, \ and\ \bibinfo {author}
  {\bibfnamefont {P.}~\bibnamefont {Werner}},\ }\href {\doibase
  10.1103/RevModPhys.83.349} {\bibfield  {journal} {\bibinfo  {journal}
  {Reviews of Modern Physics}\ }\textbf {\bibinfo {volume} {83}},\ \bibinfo
  {pages} {349} (\bibinfo {year} {2011})}\BibitemShut {NoStop}%
\bibitem [{\citenamefont {Poteryaev}\ \emph {et~al.}()\citenamefont
  {Poteryaev}, \citenamefont {Belozerov}, \citenamefont {Dyachenko},
  \citenamefont {Korotin}, \citenamefont {Korotin}, \citenamefont {Shorikov},
  \citenamefont {Skorikov}, \citenamefont {Skornyakov},\ and\ \citenamefont
  {Streltsov}}]{AMULET}%
  \BibitemOpen
  \bibfield  {author} {\bibinfo {author} {\bibfnamefont {A.}~\bibnamefont
  {Poteryaev}}, \bibinfo {author} {\bibfnamefont {A.}~\bibnamefont
  {Belozerov}}, \bibinfo {author} {\bibfnamefont {A.}~\bibnamefont
  {Dyachenko}}, \bibinfo {author} {\bibfnamefont {D.}~\bibnamefont {Korotin}},
  \bibinfo {author} {\bibfnamefont {M.}~\bibnamefont {Korotin}}, \bibinfo
  {author} {\bibfnamefont {A.}~\bibnamefont {Shorikov}}, \bibinfo {author}
  {\bibfnamefont {N.}~\bibnamefont {Skorikov}}, \bibinfo {author}
  {\bibfnamefont {S.}~\bibnamefont {Skornyakov}}, \ and\ \bibinfo {author}
  {\bibfnamefont {S.}~\bibnamefont {Streltsov}},\ }\href
  {http://amulet-code.org} {\enquote {\bibinfo {title} {{AMULET}},}\
  }\BibitemShut {NoStop}%
\bibitem [{\citenamefont {Anisimov}\ \emph {et~al.}(2009)\citenamefont
  {Anisimov}, \citenamefont {Korotin}, \citenamefont {Streltsov}, \citenamefont
  {Kozhevnikov}, \citenamefont {Kune{\v{s}}}, \citenamefont {Shorikov},\ and\
  \citenamefont {Korotin}}]{Anisimov2009a}%
  \BibitemOpen
  \bibfield  {author} {\bibinfo {author} {\bibfnamefont {V.~I.}\ \bibnamefont
  {Anisimov}}, \bibinfo {author} {\bibfnamefont {D.~M.}\ \bibnamefont
  {Korotin}}, \bibinfo {author} {\bibfnamefont {S.~V.}\ \bibnamefont
  {Streltsov}}, \bibinfo {author} {\bibfnamefont {A.~V.}\ \bibnamefont
  {Kozhevnikov}}, \bibinfo {author} {\bibfnamefont {J.}~\bibnamefont
  {Kune{\v{s}}}}, \bibinfo {author} {\bibfnamefont {A.~O.}\ \bibnamefont
  {Shorikov}}, \ and\ \bibinfo {author} {\bibfnamefont {M.~A.}\ \bibnamefont
  {Korotin}},\ }\href {\doibase 10.1134/S0021364008230069} {\bibfield
  {journal} {\bibinfo  {journal} {JETP Letters}\ }\textbf {\bibinfo {volume}
  {88}},\ \bibinfo {pages} {729} (\bibinfo {year} {2009})}\BibitemShut
  {NoStop}%
\bibitem [{\citenamefont {Dyachenko}\ \emph {et~al.}(2016)\citenamefont
  {Dyachenko}, \citenamefont {Shorikov}, \citenamefont {Lukoyanov},\ and\
  \citenamefont {Anisimov}}]{Dyachenko2016}%
  \BibitemOpen
  \bibfield  {author} {\bibinfo {author} {\bibfnamefont {A.~A.}\ \bibnamefont
  {Dyachenko}}, \bibinfo {author} {\bibfnamefont {A.~O.}\ \bibnamefont
  {Shorikov}}, \bibinfo {author} {\bibfnamefont {A.~V.}\ \bibnamefont
  {Lukoyanov}}, \ and\ \bibinfo {author} {\bibfnamefont {V.~I.}\ \bibnamefont
  {Anisimov}},\ }\href {\doibase 10.1103/PhysRevB.93.245121} {\bibfield
  {journal} {\bibinfo  {journal} {Physical Review B}\ }\textbf {\bibinfo
  {volume} {93}},\ \bibinfo {pages} {245121} (\bibinfo {year}
  {2016})}\BibitemShut {NoStop}%
\bibitem [{\citenamefont {Birch}(1947)}]{Birch1947}%
  \BibitemOpen
  \bibfield  {author} {\bibinfo {author} {\bibfnamefont {F.}~\bibnamefont
  {Birch}},\ }\href {\doibase 10.1103/PhysRev.71.809} {\bibfield  {journal}
  {\bibinfo  {journal} {Physical Review}\ }\textbf {\bibinfo {volume} {71}},\
  \bibinfo {pages} {809} (\bibinfo {year} {1947})}\BibitemShut {NoStop}%
\bibitem [{\citenamefont {Togo}\ \emph {et~al.}(2008)\citenamefont {Togo},
  \citenamefont {Oba},\ and\ \citenamefont {Tanaka}}]{Togo2008}%
  \BibitemOpen
  \bibfield  {author} {\bibinfo {author} {\bibfnamefont {A.}~\bibnamefont
  {Togo}}, \bibinfo {author} {\bibfnamefont {F.}~\bibnamefont {Oba}}, \ and\
  \bibinfo {author} {\bibfnamefont {I.}~\bibnamefont {Tanaka}},\ }\href
  {\doibase 10.1103/PhysRevB.78.134106} {\bibfield  {journal} {\bibinfo
  {journal} {Physical Review B}\ }\textbf {\bibinfo {volume} {78}},\ \bibinfo
  {pages} {134106} (\bibinfo {year} {2008})}\BibitemShut {NoStop}%
\bibitem [{\citenamefont {Togo}\ and\ \citenamefont {Tanaka}(2015)}]{Togo2015}%
  \BibitemOpen
  \bibfield  {author} {\bibinfo {author} {\bibfnamefont {A.}~\bibnamefont
  {Togo}}\ and\ \bibinfo {author} {\bibfnamefont {I.}~\bibnamefont {Tanaka}},\
  }\href {\doibase 10.1016/j.scriptamat.2015.07.021} {\bibfield  {journal}
  {\bibinfo  {journal} {Scripta Materialia}\ }\textbf {\bibinfo {volume}
  {108}},\ \bibinfo {pages} {1} (\bibinfo {year} {2015})}\BibitemShut {NoStop}%
\bibitem [{\citenamefont {Vinet}\ \emph {et~al.}(1987)\citenamefont {Vinet},
  \citenamefont {Smith}, \citenamefont {Ferrante}, \citenamefont {R},\ and\
  \citenamefont {James}}]{Vinet1987}%
  \BibitemOpen
  \bibfield  {author} {\bibinfo {author} {\bibfnamefont {P.}~\bibnamefont
  {Vinet}}, \bibinfo {author} {\bibfnamefont {J.~R.}\ \bibnamefont {Smith}},
  \bibinfo {author} {\bibfnamefont {J.}~\bibnamefont {Ferrante}}, \bibinfo
  {author} {\bibnamefont {R}}, \ and\ \bibinfo {author} {\bibfnamefont
  {H.}~\bibnamefont {James}},\ }\href {\doibase
  doi.org/10.1103/PhysRevB.35.1945} {\bibfield  {journal} {\bibinfo  {journal}
  {Physical Review B}\ }\textbf {\bibinfo {volume} {35}},\ \bibinfo {pages}
  {1945} (\bibinfo {year} {1987})}\BibitemShut {NoStop}%
\end{thebibliography}%
\end{document}